\documentclass[twocolumn]{aastex6}
\pdfoutput=1 
\usepackage[T1]{fontenc}
\usepackage{amsmath,amstext}
\usepackage{apjfonts} 
\usepackage[figure,figure*]{hypcap}
\usepackage{microtype}
\usepackage{caption}
\newcommand{\repeater}{FRB\,121102}

\shorttitle{Associating FRBs with Their Host Galaxies}
\shortauthors{Eftekhari \& Berger}

\begin{document}
\title{Associating Fast Radio Bursts with Their Host Galaxies}
\author{T. Eftekhari \& E. Berger}
\affil{Harvard-Smithsonian Center for Astrophysics, 60 Garden Street, Cambridge, Massachusetts 02138, USA}

\begin{abstract}

The first precise localization of a fast radio burst has began to shed light on the nature of these mysterious bursts and the physical mechanisms that power them. Increasing the sample of FRBs with robust host galaxy associations is the key impetus behind on-going and upcoming searches and facilities. Here, we quantify the robustness of FRB-host galaxy associations as a function of localization area and galaxy apparent magnitude. We also explore the use of FRB dispersion measures to constrain the source redshift, thereby reducing the number of candidate hosts. We use these results to demonstrate that even in the absence of a unique association, a constraint can be placed on the maximum luminosity of a host galaxy as a function of localization and DM. We find that localizations of $\lesssim 0.5''$ are required for a chance coincidence probability of $\lesssim 1\%$ for dwarf galaxies at $z\gtrsim 0.1$; if some hosts have luminosities of $\sim L^*$, then localizations of up to $\approx 5''$ may suffice at $z\sim 0.1$. Constraints on the redshift from the DM only marginally improve the association probability, unless the DM is low, $\lesssim 400$ pc cm$^{-3}$. This approach also relies on the determination of galaxy redshifts, which is challenging at $z\gtrsim 0.5$ if the hosts are dwarf galaxies. Finally, interesting limits on the maximum host luminosity require localizations of $\lesssim 5''$ at $z\gtrsim 0.1$. Even a few such localizations will shed light on the nature of FRB progenitors, their possible diversity, and their use as cosmological tools. 

\end{abstract}

\keywords{radio continuum: transients -- galaxies: distances and redshifts -- methods: statistical}

\section{Introduction}\label{sec:intro}

Fast radio bursts (FRBs) are bright millisecond-duration GHz frequency flares with dispersion measures well in excess of Galactic values. The high brightness temperatures and short durations point to coherent radio emission emerging from a compact source. The first FRB was discovered in archival data by \citet{Lorimer2007} and subsequently about 20 additional FRBs have been found in both archival and real-time searches (\citealt{Keane2012}; \citealt{Thornton2013}; \citealt{Spitler2014}; \citealt{Petroff2016}; \citealt{Champion2016}; \citealt{Caleb2017}). The poor localizations of FRBs ($\sim 10^3$ arcmin$^2$) have allowed for wide speculation about their distance scale and possible progenitor systems, including giant pulses from young pulsars (\citealt{Cordes2016}; \citealt{Lyutikov2016}), the collapse of supramassive neutron stars to black holes \citep{Falcke2014}, magnetar flares (\citealt{Popov2013}; \citealt{Lyubarsky2014}); mergers of compact objects (\citealt{Zhang2016}; \citealt{Wang2016}), and emission from rapidly spinning magnetars (\citealt{Metzger2017}; \citealt{Kumar2017}). A number of Galactic models have also been proposed (\citealt{Bannister2014}; \citealt{Maoz2015}).

To distinguish between the various models and to make progress in the understanding of FRBs requires more precise localizations. These can be achieved in two ways. First, via the detections of longer duration counterparts\footnote{We note that in general $\gamma$-ray localizations will not substantially improve on the FRB single-dish localizations.} in the radio, NIR/optical/UV, or X-rays (referred to here as ``indirect localization''). Second, via the detection of FRBs with radio interferometers, which can provide much more precise positions than single-dish radio telescopes (referred to here as ``direct localization''). Undertaking the indirect approach, \citet{Keane2016} claimed the first precise localization of an FRB based on the detection of a long-lived radio counterpart coincident with the nucleus of a galaxy at $z\approx 0.49$ within the Parkes Telescope localization region. However, the claimed counterpart was subsequently shown to be a variable AGN and hence unrelated to the FRB \citep{Williams2016}. To date, no other transient counterparts have been claimed for FRBs despite some real-time follow-up efforts \citep{Petroff2015}.

The discovery of the repeating \repeater\ \citep{Spitler2014,Spitler2016} has successfully enabled the second approach of direct localization.  The detection of repeating bursts from this source with the Karl G.~Jansky Very Large Array (VLA) and the European Very Long Baseline Interferometry Network (EVN) has led to a sub-arcsecond localization \citep{Chatterjee2017}, as well as the discovery of an associated quiescent radio source \citep{Marcote2017}. Subsequent optical observations revealed a coincident low metallicity star forming dwarf galaxy at $z=0.193$ \citep{Tendulkar2017}. 

The existence of the repeating \repeater\ argues against cataclysmic formation scenarios, while the association with an extragalactic source rules out a Galactic origin. Indeed, the properties of \repeater, its host galaxy, and the associated quiescent radio source are consistent with a model in which the FRBs are powered by a millisecond magnetar formed in a superluminous supernova (SLSN) or a long-duration gamma-ray burst (LGRB) a few decades ago (\citealt{Metzger2017}; see also \citealt{Piro2016}; \citealt{Murase2016}). Recent work by \citet{Nicholl2017} demonstrates that this scenario can be extended to the broad FRB population if the typical source lifetimes are $\sim 10^2$ years, and that in general the formation channel can be robustly tested with $\lesssim 10$ additional FRB localizations. A number of upcoming radio facilities and dedicated surveys aim to detect large numbers of FRBs and to asses their repeatability. However, these facilities will have a wide range of localization capabilities, ranging from $\lesssim 1''$ to $\sim 10'$ or even larger. 

Motivated by FRB 121102, its extragalactic origin, and its association with a galaxy, we assess how robustly similar host galaxy associations can be made, assuming that FRBs are associated with common constituents of galaxies. We explore these associations based on the localization capabilities of various facilities and the properties of the galaxies, making no a priori assumptions about the nature of the host galaxies. We also explore whether redshift information as inferred from the dispersion measure can improve the association confidence.  Finally, we consider what limits can be placed on the host galaxy luminosities even in the absence of a robust host association. Our results are assessed against a wide range of FRB search facilities and galaxy luminosities that can be expected in various FRB models. We stress that while we focus on localization via the direct method, our results hold for any FRBs localized indirectly via emission in other bands (radio to $\gamma$-rays and non-electromagnetic). 

The structure of the paper is as follows.  In \S\ref{sec:nc} we utilize deep optical galaxy number counts spanning to $\approx 30$ mag to determine the probability of chance coincidence for an FRB and a potential host galaxy as a function of localization area and galaxy brightness.  In \S\ref{sec:dm} we additionally incorporate limits on the redshift based on DM information, taking into account uncertainty in the intergalactic medium DM-redshift relation.  In \S\ref{sec:ul} we demonstrate that even in the absence of a robust host association, interesting limits can be placed on the host galaxy luminosity, for sufficiently well-localized events.  We discuss our results in the context of various FRB facilities in \S\ref{sec:disc}.

\section{Assessing Probability of Chance Coincidence with Galaxy Number Counts}\label{sec:nc}

\begin{figure*}
\includegraphics[width=\textwidth]{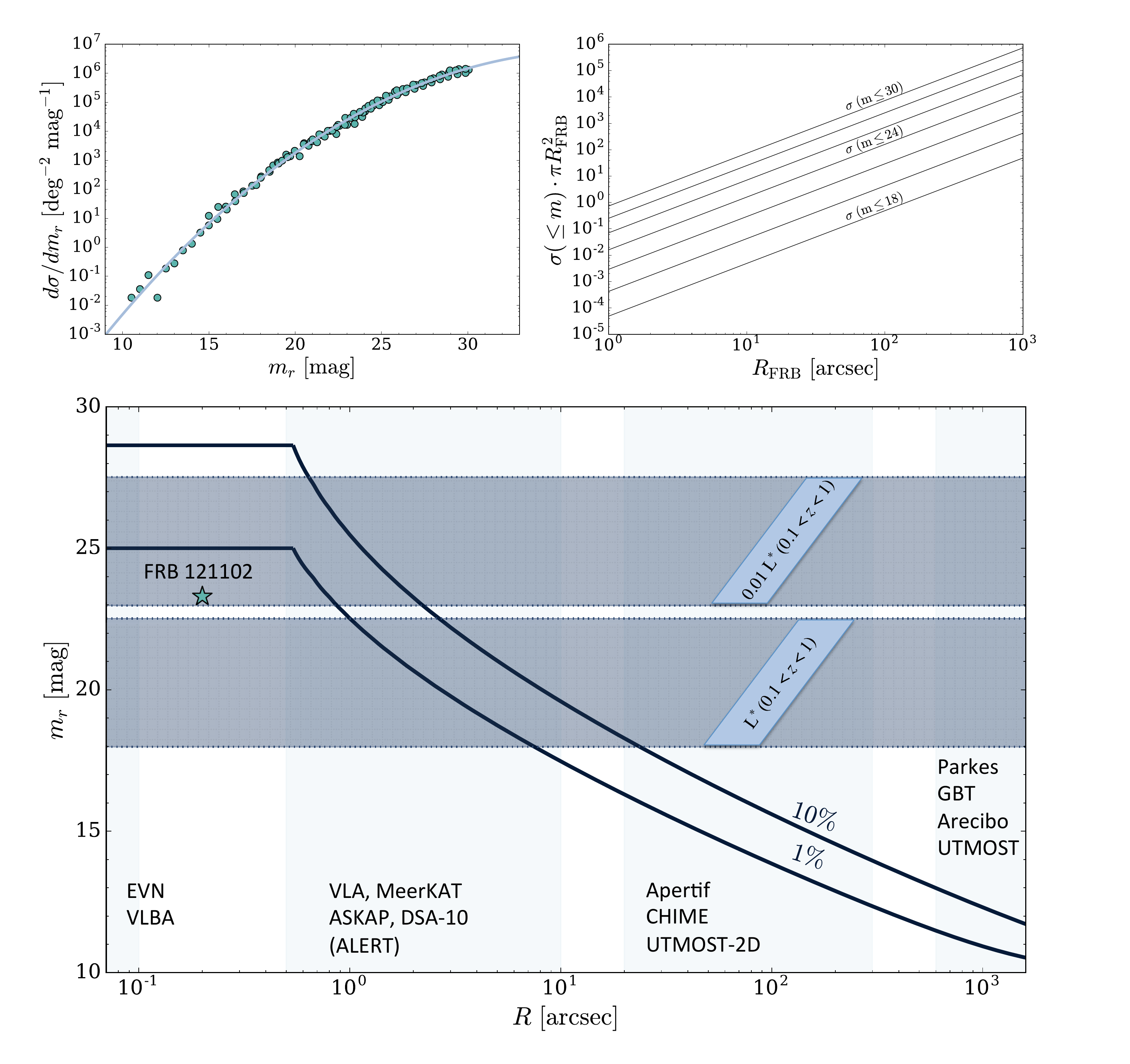}
\caption{\textit{Top-Left:} Galaxy number counts in $r$-band \citep{Driver2016}, along with our spline fit to the data (cyan line). \textit{Top-Right:} The number of galaxies brighter than $m_r$ ($\sigma(\leq m)\pi R^2_{\rm FRB})$) as a function of FRB localization radius ($R_{\rm FRB}$), as derived from integration of the spline fit to the number counts in the top-left panel. Individual lines correspond to increments of 2 mag, spanning from 18 to 30 mag; the bright end corresponds to an $L^*$ galaxy at $z\approx 0.1$.  \textit{Bottom:} Probability contours for $P_{\rm cc}=0.01$ and 0.1 as a function of $r$-band magnitude and radius ($R$). The plateau below $\approx 0.6''$ corresponds to the regime in which the radius is set by the FRB-galaxy offset or the galaxy half-light radius; at larger radii $R=2 R_{\rm FRB}$. Horizontal bands indicate the luminosities of 0.01 $L^*$ and $L^*$ galaxies at $z=0.1-1$.  We also plot the data for the localization and host of \repeater\ (corrected for Galactic extinction; \citealt{Chatterjee2017,Marcote2017,Tendulkar2017}). Vertical bars indicate the localization regimes of various radio telescopes that are designed to or capable of detecting FRBs (Table 1). Only facilities capable of sub-arcsecond localization can lead to robust associations with $0.01$ $L^*$ host galaxies. If some hosts are $L^*$ galaxies, then localizations of a few arcseconds may suffice.}
\label{fig:nc}
\end{figure*}

We assess the association of an FRB and a putative host galaxy by calculating the probability of chance coincidence for a galaxy of a given brightness in the FRB localization region. We use a spline fit to the $r$-band galaxy number counts presented in \citet{Driver2016} to calculate the projected areal number density of galaxies brighter than $r$-band magnitude, $\sigma(\leq m)$; see top left panel of Figure~\ref{fig:nc}. We note that the results are insensitive to the choice of filter within the optical regime. Assuming a Poisson distribution of galaxies across the sky, the probability of a chance coincidence occurring within a radius $R$ is given by: 
\begin{equation} 
P_{\rm cc} = 1 - e^{-\pi R^2 \sigma(\leq m)}.
 \end{equation}
Following the prescription of \citet{Bloom2002}, we parameterize the localization region using $R={\rm max}[2R_{\rm FRB}, \sqrt{R_0^2 + 4R_h^2}]$, where $R_{\rm FRB}$ is the $1\sigma$ localization radius of the FRB (or equivalently, the radius that corresponds to $1\sigma$ of the localization area), $R_0$ is the radial angular separation between the FRB position and a presumed host, and $R_h$ is the galaxy half-light radius.  For a poor localization with a galaxy inside the localization region, the first term dominates. Conversely, $\sqrt{R_0^2 + 4R_h^2}$ describes the regime in which the FRB localization is precise, but the dominant scale is the angular size of the galaxy or the offset between the FRB and the galaxy.

\begin{deluxetable}{lccc}
\tablecolumns{4}
\tablewidth{0pt}  
\caption{Radio facilities and their localization capabilities.}
\tablehead{
\colhead{Telescope} & 
\colhead{$R_{\rm FRB}$} & 
\colhead{$m_r(P_{\rm cc} = 0.01)$} & 
\colhead{$m_r(P_{\rm cc} = 0.1)$} \\ 
\colhead{} & 
\colhead{[arcsec]} & 
\colhead{[mag]} & 
\colhead{[mag]} 
}  
\startdata
VLBA / EVN & $0.001\text{--}0.1$ & 25 & 28 \\
VLA & $0.1\text{--}3$ & 21 & 24 \\
ASKAP & $0.8\text{--}1.5$ & 21 & 23 \\
DSA-10 & $1\text{--}2$ & 20 & 22 \\
MeerKAT & $2\text{--}10$ & 17 & 19 \\
UTMOST-2D & $2\text{--}30$ & 16 & 18 \\
Apertif$^{\dagger}$ & $5\text{--}60$ & 14 & 17 \\
CHIME & $20\text{--}600$ & 14 & 15 \\
UTMOST & $100\text{--}300$ & 13 & 15 \\
Arecibo & $200\text{--}230$ & 12 & 14 \\
Parkes & $500\text{--}800$ & 11 & 13 \\
GBT & $500\text{--}800$ & 11 & 13 \\
\enddata
\tablecomments{Radio facilities capable of detecting FRBs, ordered by localization capability. $\dagger$ In conjunction with LOFAR, the Apertif LOFAR Exploration of the Radio Transient Sky (ALERT)\footnote{\citealt{vanLeeuwen2014}; see also http://alert.eu/.} survey can provide more accurate (arcsecond) localizations.}
\end{deluxetable}

Since FRB\,121102 was localized to a dwarf galaxy, and SLSNe and LGRBs are associated with similar hosts, we use the results from \citet{Blanchard2016} and \citet{Lunnan2015} to characterize $R_0$ and $R_h$ for LGRB and SLSN host galaxies, respectively. The median galaxy offsets and half-light radii for LGRBs at $z\lesssim 1$ are $R_0\approx 0.2''$ and $R_h \approx 0.3''$, respectively.  Similarly, for SLSNe, they are $R_0\approx 0.2''$ and $R_h\approx 0.2''$.  We therefore adopt characteristic values of $R_0 = 0.2''$ and $R_h=0.25''$.  Thus, for $R_{\rm FRB}\lesssim 0.3''$, the relevant scales are the galaxy half-light radius or offset, and not the localization radius.  This result holds for dwarf galaxies in general, which have half-light radii of $\approx 1-2$ kpc (e.g., \citealt{shen2003}), or correspondingly $\approx 0.5-1''$ at $z\approx 0.1$ and $\approx 0.12-0.25''$ at $z\approx 1$.  We note that if FRBs turn out to generally originate in more luminous (and hence larger) galaxies, then the transition between the two regimes will occur at a larger scale, $R_{\rm FRB}\sim {\rm few}$ arcsec.

Using the results from the galaxy number count integration, we plot the total number of galaxies above some limiting magnitude [i.e., $\pi R_{\rm FRB}^2\sigma(\leq m$)] as a function of localization size in the top right panel of Figure~\ref{fig:nc}.  The plot shows the results for the range of $18-30$ mag, with the bright end corresponding to an $L^*$ galaxy at $z\approx 0.1$.  We find that at even at the bright end ($18$ mag) the expectation value is about 1 galaxy per $R_{\rm FRB}\approx 150''$ radius region, which is about 30 times smaller than the typical localization area of existing FRBs from single-dish telescopes.  This clearly demonstrates why finer localizations are required for host associations.  At the faint end ($30$ mag) the expectation value is about 1 galaxy per $R_{\rm FRB}\approx 1''$ region.  We note that at the expectation value of 1 galaxy the resulting probability of chance coincidence is large, $P_{\rm cc}\approx 0.6$, so this would not lead to a robust association (we return to this point in \S\ref{sec:ul}).

In the lower panel of Figure~\ref{fig:nc}, we plot the resulting probability contours for chance coincidence as a function of apparent $r$-band magnitude and localization region ($R$). We specifically show the $P_{\rm cc}=0.01$ and $0.1$ contours, which delineate a reasonable range for robust associations. Also shown are the apparent magnitude bands corresponding to 0.01 $L^*$ and $L^*$ galaxies in the redshift range $z\approx 0.1-1$ using the galaxy luminosity functions of \citet{Blanton2003} and \citet{Beare2015}; see \S\ref{sec:dm}. The plateau in the probability contours below $R\approx 0.6''$ illustrates the transition to the regime in which $R$ is set by the FRB-galaxy angular offset or the galaxy half-light radius. 

At $z\approx 0.1$ and $\approx 1$, a galaxy with 0.01 $L^*$ has an apparent magnitude of $m_r\approx 22.5$ and $\approx 27.5$, respectively. For a 0.01 $L^*$ galaxy at $z\gtrsim 0.1$, sub-arcsecond ($R_{\rm FRB}\lesssim 0.4''$) localizations are required for chance coincidence probabilities of $P_{\rm cc}\lesssim 0.01$. Finer localizations do not improve the association confidence because of the transition to the offset or galaxy size dominated regime.  Due to the same effect, at $z\gtrsim 0.5$ the chance coincidence probability will be limited to $P_{\rm cc}\gtrsim 0.01$ for 0.01 $L^*$ galaxies. An $L^*$ galaxy at $z\gtrsim 0.1$ requires a localization of $R_{\rm FRB}\lesssim 3''$ for $P_{\rm cc}\approx 0.01$, while at $z\approx 1$ an $R_{\rm FRB}\lesssim 0.5''$ localization is required. For $R\gtrsim 30''$ ($R_{\rm FRB}\gtrsim 15''$), associations with $P_{\rm cc}\lesssim 0.1$ are impractical since they require a galaxy several magnitudes more luminous than $L^*$ at a small redshift of $z\lesssim 0.1$.

In Table 1, we list existing radio facilities capable of or designed to detect FRBs. These are sorted by anticipated or known localization capability. Vertical bands in the lower panel of Figure~\ref{fig:nc} depict the expected localization regimes for the various facilities. We also include in Table 1 the apparent magnitude of a galaxy that would result in a 1\% and 10\% chance coincidence probability, based on the respective localization capabilities, where the values are derived directly from the probability contours in Figure~\ref{fig:nc} (see \S\ref{sec:disc} for a detailed discussion). 

\begin{figure*}
\includegraphics[width=\textwidth]{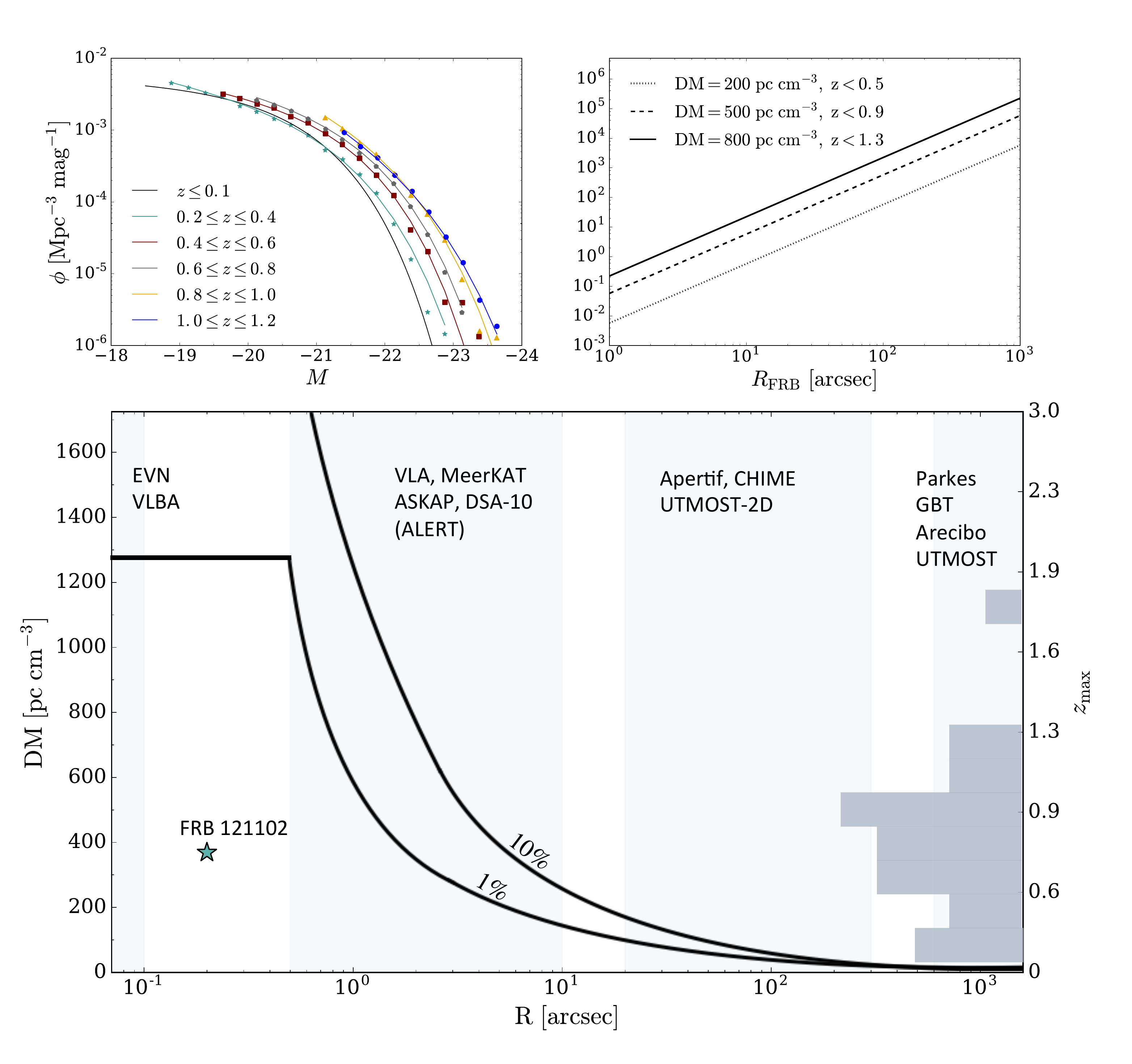}
\caption{\textit{Top-Left:} Luminosity functions for blue galaxies out to $z\sim 1.2$ from \citet{Blanton2003} and \citet{Beare2015}, using a cosmology with $\Omega_0=0.3$, $\Omega_\Lambda=0.7$, and $H_0=70$ km s$^{-1}$ Mpc$^{-1}$. \textit{Top-Right:} The number of galaxies with $\gtrsim 0.01$ $L^*$ out to a redshift $z_{\rm max}$ (for representative values of DM and taking into account the scatter about the mean DM($z$) relation; see \S\ref{sec:dm}) as a function of localization radius. \textit{Bottom:} Probability contours for a $P_{\rm cc} = 0.01$ and $0.1$ as a function of DM and localization radius. Also shown as a histogram along the right axis is the maximum redshift distribution of known FRBs, after subtraction of the Milky Way contribution \citep{Petroff2016}.  We also plot the data for \repeater\ (with the Galactic contribution to the DM value subtracted off; \citealt{Chatterjee2017}).  Vertical bars indicate the localization regimes of various radio telescopes that are designed to or capable of detecting FRBs (Table 1). Associations with $P_{\rm cc}=0.01$ are limited to cases of ${\rm DM}\lesssim 1100$, while at low DM values of $\lesssim 200$ pc cm$^{-3}$, a robust association can tolerate a larger localization region.}
\label{fig:dm}
\end{figure*}

\section{Utilizing Redshift Constraints from the Dispersion Measure}\label{sec:dm}

In the case of FRBs, we can use the measured DM to constrain the source redshift. This effectively reduces the number of chance coincidences by eliminating galaxies that contribute to the number counts but are located beyond the redshift range of interest. To assess the fractional gain in association probability when incorporating redshift information, we calculate the number density of galaxies by integrating the optical luminosity functions presented in \citet{Beare2015} and \citet{Blanton2003}, which cover $z\approx 0.2-1.2$ and $z\approx 0.1$, respectively; see top-right panel of Figure~\ref{fig:dm}. Motivated by the host of FRB\,121102, we utilize the luminosity functions for blue galaxies.  We note that the majority of FRBs detected at the time of this publication have ${\rm DM}\lesssim 1200$ pc cm$^{-3}$, corresponding to  $z\lesssim 1.3$ using the redshift-DM relation of \citep{Deng2014} (see also \citealt{Ioka2003}).

The total number of galaxies above a certain luminosity out to some redshift is obtained by multiplying the galaxy volume density by the comoving volume out to the relevant redshift and the fractional area of the localization region on the sky $f_A=\pi R^2/5.346\times 10^{11}$, where $R$ is in units of arcsec. The probability of chance coincidence is then given by: 
\begin{equation}
P_{\rm cc} = 1 - e^{-f_A \Sigma(\leq M) V_{C}(\leq z)},
\end{equation}
where $\Sigma(\leq M)$ is the number density of galaxies obtained from integrating the luminosity functions above a limiting absolute magnitude $M$, and $V_C(\leq z)$ is the comoving volume out to a redshift $z$. 

To account for inhomogeneities in the distribution of baryons across varying sight lines, we utilize the uncertainty in DM($z$) as derived in \citet{McQuinn2014}. In that analysis, the largest scatter about the mean DM($z$) relation occurs when baryons trace dark matter halos, assuming NFW profiles. We incorporate this uncertainty, which provides the most conservative limit, in our estimate of the redshift upper limit, $z_{\rm max}$. For example, for ${\rm DM}\approx 800$ pc cm$^{-3}$, the standard deviation about the mean DM($z$) relation is $\approx 400$ pc cm$^{-3}$, corresponding in turn to an upper limit of $z_{\rm max}\approx 1.3$. At the low range for FRBs, ${\rm DM}\approx 200$ pc cm$^{-3}$, the uncertainty in the DM($z$) relation corresponds to $z_{\rm max}\approx 0.5$. For a given DM value we therefore determine $z_{\rm max}$, and carry out the volume integration in redshift annuli that correspond to the luminosity function redshift bins shown in the top-left panel of Figure~\ref{fig:dm}.  In the top-right panel of Figure~\ref{fig:dm} we show the resulting number of galaxies as a function of localization radius for representative DM values of 200, 500, and 800 pc cm$^{-3}$. 

In principle, the standard deviation about the mean DM($z$) relation can also be used to place a lower limit on the  redshift.  However, we do not take this approach here since contributions to the DM from the host galaxy and/or local environment of the FRB are likely, but not generally known.  Such contributions will reduce the lower bound on the redshift further, leading to a significant uncertainty in the actual allowed lower redshift.  Since the comoving volume is substantially smaller at low redshift, the absence of a lower bound does not change the results significantly. 

We find that at large DM values ($\approx 800$ pc cm$^{-3}$; $z_{\rm max}\approx 1.3$) the use of DM only affords a factor of two fewer galaxies compared to the use of number counts alone (\S\ref{sec:nc}). However, for the lower end of FRB DM values, corresponding to $z_{\rm max}\approx 0.5$, this provides about an order of magnitude fewer galaxies within a given localization region, and hence allows for a robust FRB-host association with a poorer localization region compared to the use of number counts alone.

In Figure~\ref{fig:dm}, we also plot the $1\%$ and $10\%$ chance coincidence probability contours as a function of DM and localization radius, where the contours are derived using Equation 2 for galaxies with $\ge 0.01$ $L^*$. The contours indicate that robust associations at higher DMs (larger $z_{\rm max}$) require finer localizations in order to compensate for the larger cosmic volume. A value of $P_{\rm cc}=0.01$ for ${\rm DM}\approx 1000$ pc cm$^{-3}$ requires sub-arcsecond localizations, while associations of comparable probability can be made for larger localizations ($\approx 1''$) when ${\rm DM}\lesssim 400$ pc cm$^{-3}$. However, even for the lowest measured DM values in the current FRB sample,  localizations of $R_{\rm FRB}\gtrsim 5''$ lead to poor association probability. As a specific example, at the DM of \repeater, even a localization with $R_{\rm FRB}\approx 1''$ would provide an association with $P_{\rm cc}\approx 0.01$ when using the relevant redshift upper limit (i.e., the very precise localization of \repeater, $R_{\rm FRB}\lesssim 0.1''$ is an over-kill).  Figure~\ref{fig:dm} shows the redshift distribution of $z_{\rm max}$ for known FRBs\footnote{http://www.astronomy.swin.edu.au/pulsar/frbcat/} \citep{Petroff2016} for the purpose of comparison. The localization capabilities of various facilities are also shown; these are discussed in detail in \S\ref{sec:disc}.

\begin{figure*}
\includegraphics[width=\textwidth]{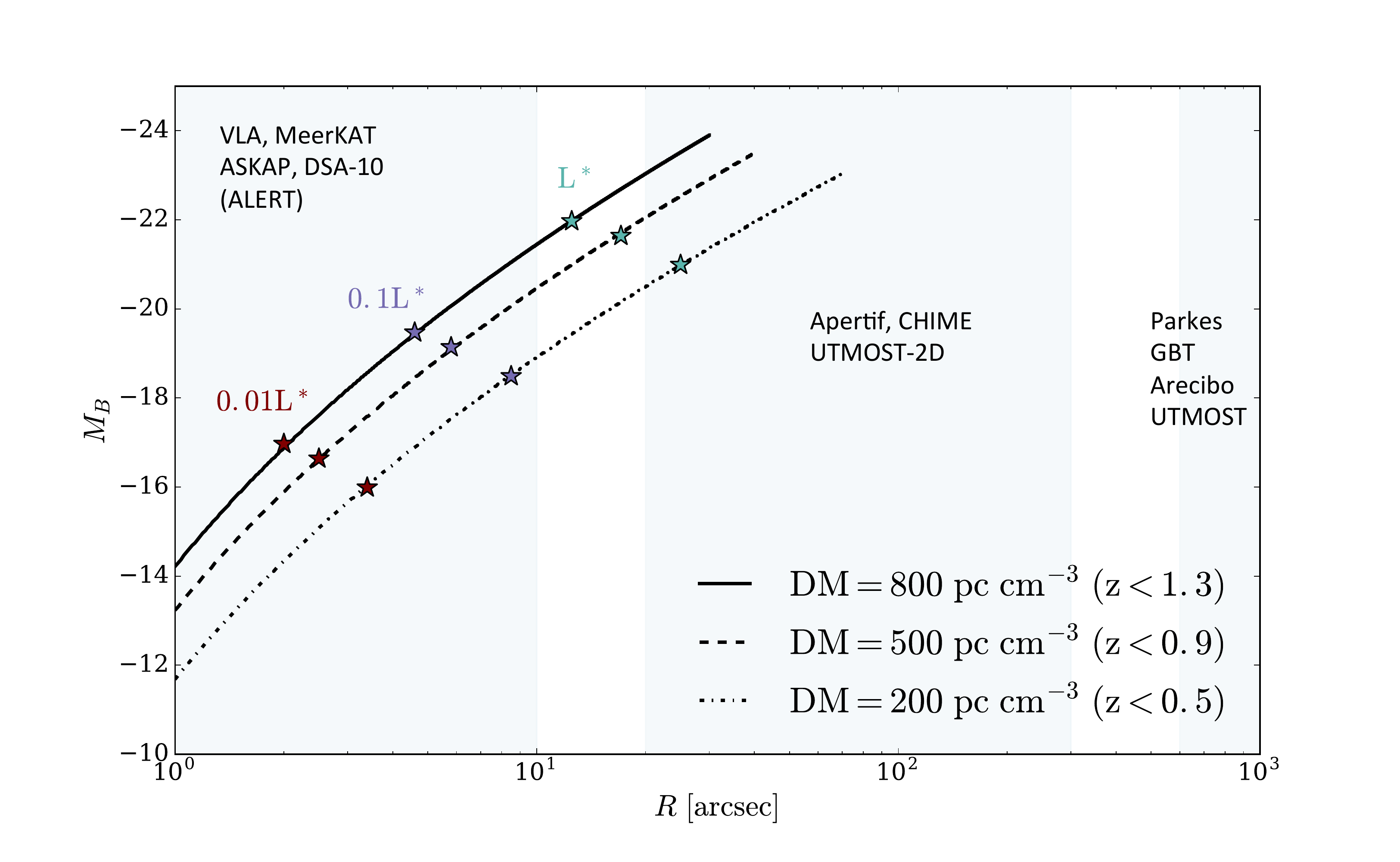}
\caption{The upper limit on the absolute magnitude of a host galaxy as a function of localization radius for a range of DM values.  The limit is calculated using the number counts in Figure~\ref{fig:nc} set to an expectation value of 1 galaxy, and the upper limit on the redshift for each DM value. Individual curves are truncated around the maximum expected luminosity for galaxies at the associated redshift ($\sim 5$ $L^*$). Also shown are the $0.01$ $L^*$, $0.1$ $L^*$, and $L^*$ values for each DM value.  Vertical bars indicate the localization regimes of various radio telescopes that are designed to or capable of detecting FRBs (Table 1).  Meaningful constraints on the host galaxy luminosity (e.g., separating $L^*$ and dwarf galaxies) require positions of $\lesssim 5''$.}
\label{fig:ul}
\end{figure*}

\section{Placing an Upper Limit on the Host Galaxy Luminosity}\label{sec:ul}

In addition to estimating the robustness of a unique host galaxy association, we can use the results of our calculations to estimate an upper limit on the luminosity of the host galaxy even in the absence of a robust association.  Using the galaxy number counts results from \S\ref{sec:nc}, we can determine, as a function of localization region, the apparent magnitude associated with a number density that would lead to an expected 1 galaxy within an FRB localization region. This provides an estimate of the brightest expected galaxy; fainter galaxies will be more numerous.  We note that at this expectation value the probability of chance coincidence is by definition large, $P_{\rm cc}\approx 0.6$ so it precludes a robust association.  As in \S\ref{sec:dm}, we can then use the DM value to infer $z_{\rm max}$ and thereby calculate the corresponding luminosity of the galaxy at that redshift. Since we combine the brightest apparent magnitude with the maximum redshift, this given an absolute upper limit on the host luminosity.  Figure~\ref{fig:ul} shows the expected maximum absolute magnitude of a host galaxy as a function of localization radius for a range of DM values. Individual curves are truncated at 5 $L^*$, corresponding to the most luminous galaxies.  As in Figures~\ref{fig:nc} and \ref{fig:dm}, we overlay radio facility localization capabilities for reference. 

The results suggest that for ${\rm DM}\approx 800$ pc cm$^{-3}$ a meaningful upper limit on the luminosity of the host galaxy (i.e., $\lesssim 5$ $L^*$) can be made only if $R_{\rm FRB}\lesssim 15''$; for ${\rm DM}\approx 200$ pc cm$^{-3}$, localizations of $R_{\rm FRB}\lesssim 30''$ can be used to place a similar constraint.  However, to place an interesting upper limit on the host luminosity, for example $\lesssim 0.1$ $L^*$ to determine if the host is a dwarf galaxy, requires localizations of $R_{\rm FRB}\lesssim 3''$ at ${\rm DM}\approx 800$ pc cm$^{-3}$; an upper limit of $\lesssim 0.01$ $L^*$ requires a localization of $R_{\rm FRB}\lesssim 1''$.  Thus, even a  constraint on the host galaxy luminosity (albeit without a robust association) requires localizations of at most a few arcseconds.

\section{Discussion}\label{sec:disc}

The information summarized in Figure~\ref{fig:nc} indicates that an FRB-host association with $P_{\rm cc}\lesssim 0.01$ requires a host galaxy brighter than $m_r\approx 25$ mag and a sub-arcsecond localization. This is the case for \repeater, which was localized to $\lesssim 0.1''$ \citep{Chatterjee2017,Marcote2017} and whose host galaxy (corrected for Galactic extinction) has $m_r\approx 23.3$ mag \citep{Tendulkar2017}; see Figure~\ref{fig:nc}.  The brightness level of $m_r\lesssim 25$ mag corresponds to a 0.01 $L^*$ galaxy at $z\lesssim 0.5$ or an $L^*$ galaxy at $z\lesssim 1.5$.  A number of facilities, including the VLBA, EVN, VLA, ASKAP, MeerKAT,  and DSA-10 will be capable of providing such localizations.  Although some of these facilities can provide localizations at the level of $\sim 0.1''$ or better, these will not improve the association probability markedly due to the typical angular extent of the galaxies. This is the case for FRB\,121102 \citep{Chatterjee2017,Marcote2017}. For localizations of $\approx 10''$, (e.g., at the edge of capability for CHIME, Apertif, and UTMOST-2D) even  associations with a marginal confidence of $P_{\rm cc}\approx 0.1$ can only be made for $L^*$ galaxies at $z\sim 0.1$. Larger localization regions of $\gtrsim 30''$ will not provide meaningful associations unless some FRBs are located in the most luminous galaxies at $z\lesssim 0.1$.

Thus, if FRBs originate exclusively in dwarf galaxies, confident associations with $P_{\rm cc}\lesssim 0.01$ will be out of reach of radio facilities with localization capabilities of $\gtrsim 1''$.  For localizations of $1-2''$ dwarf galaxy associations can be made at a level of $P_{\rm cc}\approx 0.1$, while for $\gtrsim 2''$ robust associations with dwarf galaxies become untenable.

The DM of the FRB can in principle be leveraged to further constrain a host association, since the galaxy number counts alone are insensitive to redshift information. The maximum redshift as derived from the dispersion measure can be used in conjunction with the luminosity functions to exclude contributions from higher redshifts (which appear in the number counts).  The gain is minimal at ${\rm DM}\sim 800$ pc cm$^{-3}$ (a factor of two fewer galaxies relative to the number counts alone), but it is more substantial at lower DM values (i.e., lower redshifts), with about an order of magnitude fewer galaxies for ${\rm DM}\sim 200$ pc cm$^{-3}$, and hence resulting localization regions that can be correspondingly larger. 

Using the dispersion measure to constrain the redshift may prove largely inconsequential, however, as the galaxies of interest may be too faint for spectroscopic and even photometric redshift determinations. The limiting magnitude for spectroscopy is $\sim 25$ mag, corresponding to a $0.01L^*$ galaxy at $z\sim 0.5$.  Thus, if FRBs originate in dwarf galaxies, obtaining spectra for their hosts at $z\gtrsim 0.5$ (i.e, ${\rm DM}\gtrsim 400$ pc cm$^{-3}$) will be beyond the limits of existing ground-based telescopes. Photometric redshifts can be obtained with the largest ground-based telescopes or with the {\it Hubble Space Telescope}, but even these will prove challenging for dwarf galaxies at $z\sim 1$, whose apparent magnitudes are $\gtrsim 27$ (Figure~\ref{fig:nc}). Thus, we conclude that the utility of dispersion measure constraints on the redshift is limited to cases of modest DM values.  Based on the existing DM distribution (Figure~\ref{fig:dm}) this accounts for at most half of the sample.  

In cases where a host galaxy cannot be robustly identified, upper limits on the luminosity of the host can be obtained using $z_{\rm max}$ as derived from the dispersion measure. For $L^*$ galaxies, this requires localizations of $R_{\rm FRB}\lesssim 8''$ for ${\rm DM}\approx 800$ pc cm$^{-3}$ ($z\sim 1.3$) and $\lesssim 15''$ for ${\rm DM}\approx 200$ pc cm$^{-3}$ ($z\sim 0.5$).  However, for 0.1 $L*$ galaxies, localization accuracies of $R_{\rm FRB}\lesssim 3''$ and $\lesssim 5''$ are required for the same DM values.  Such upper limits are required to support a connection to SLSNe and LGRBs.  In particular, \citet{Nicholl2017} demonstrate that $\lesssim 10$ host galaxy identifications with $\lesssim 0.5$ $L^*$ will be sufficient to rule out a number of FRB formation channels, including short gamma-ray bursts or any channels that trace the cosmic stellar mass or star formation rate density. Thus, a few localizations with $\lesssim 5''$ will be sufficient to place general constraints on the FRB formation channel.

\section{Conclusions}\label{sec:conc}

We assessed the level of robustness with which an FRB-host association may be made as a function of apparent magnitude and localization region, assuming that FRBs are extragalactic in origin and associated with common constituents of galaxies, as motivated by the repeating FRB 121102. We also explored the use of DM to place an upper bound on the host redshift, and the general limits that can be placed on a host luminosity even in the absence of a robust unique association.  Our key results are summarized as follows:

\begin{itemize}

\item{Sub-arcsecond localizations, $R_{\rm FRB}\lesssim 0.4''$, are required for robust host galaxy associations if the hosts are dwarf galaxies at $z\gtrsim 0.1$.  If the hosts are instead generally $L^*$ galaxies, then localizations of $\approx 3''$ may suffice at $z\approx 0.1$ (but $\approx 0.5''$ at $z\approx 1$). Several radio facilities, including the VLBA, EVN, VLA, ASKAP, DSA-10, and MeerKAT are capable of providing such localizations.  We note that localizations of $R_{\rm FRB} \lesssim 0.3''$ do not generally increase the association probability given the typical sizes of dwarf galaxies.  Facilities that provide localizations of $R_{\rm FRB}\gtrsim 10''$ will not lead to robust host associations, unless some of the hosts are the most luminous known galaxies at $z\lesssim 0.1$.}

\item{For moderate dispersion measures of $\lesssim 400$ pc cm$^{-3}$, the resulting upper limit on the source redshift can be used to increase the confidence of an association, and may therefore tolerate larger localizations of up to $\approx 1''$.  However, for larger DM values (and hence redshifts, $z\gtrsim 0.5$) the gain is minimal, and furthermore the determination of galaxy redshifts via spectroscopic or photometric methods will be challenging if the hosts are generally dwarf galaxies.}

\item{In the absence of a robust association, upper limits on the luminosity of the host galaxy can still be obtained.  However, for these constraints to be meaningful in terms of insight into FRB formation channels (e.g., \citealt{Nicholl2017}) localizations of $R_{\rm FRB}\lesssim 5''$ (at ${\rm DM}\approx 200$ pc cm$^{-3}$) and $\lesssim 3''$ (at ${\rm DM}\approx 800$ pc cm$^{-3}$) are still required.}

\end{itemize}

If FRBs are empirically shown to be associated with galaxies, increasing the number of FRB-host galaxy associations will be essential both for our understanding of FRB progenitors and their formation channels, as well as for enabling their use as cosmological probes. If \repeater\ and its host galaxy are representative of the population as a whole, localizations of $\lesssim 1''$ will be required.  Luckily, this is within the reach of the direct localization capabilities of some existing and upcoming facilities.  The key challenge for these facilities may instead be the large number of hours required to detect even a single FRB; for example, it required about 80 hours of VLA time to localize \repeater\ \citep{Chatterjee2017}. 

\acknowledgments \textit{Acknowledgments.} We thank P.~K.~G.~Williams and M.~Nicholl for helpful discussions and comments. The Berger Time-Domain Group is supported in part by NSF grant AST-1411763 and NASA ADA grant NNX15AE50G. This research has made use of the FRB Catalogue ({\tt \footnotesize{http://www.astronomy.swin.edu.au/pulsar/frbcat/}}) and NASA's Astrophysics Data System.

\bibliography{ref}
\bibliographystyle{apj}

\end{document}